\documentclass[
reprint,
superscriptaddress,
amsmath,amssymb,
aps,
prl,
]{revtex4-1}

\usepackage{graphicx}
\usepackage{dcolumn}
\usepackage{bm}
\usepackage{mhchem} 
\usepackage{amsmath} 
\usepackage{color} 
\usepackage{soul} 
\usepackage{wasysym}

\begin{document}

\preprint{APS/123-QED}

\title{Nonlinear magnetization dynamics driven by strong terahertz fields}

\author{Matthias Hudl}
\affiliation{Department of Physics, Stockholm University, 106 91 Stockholm, Sweden}

\author{Massimiliano d'Aquino}
\affiliation{Department of Engineering, University of Naples ``Parthenope'', 80143 Naples, Italy}

\author{Matteo Pancaldi}
\affiliation{Department of Physics, Stockholm University, 106 91 Stockholm, Sweden}

\author{See-Hun Yang}
\affiliation{IBM Almaden Research Center, San Jose CA 95120, USA}

\author{Mahesh G. Samant}
\affiliation{IBM Almaden Research Center, San Jose CA 95120, USA}

\author{Stuart S. P. Parkin}
\affiliation{IBM Almaden Research Center, San Jose CA 95120, USA}
\affiliation{Max-Planck Institut f\"ur Mikrostrukturphysik, 06120 Halle, Germany}

\author{Hermann A. D\"urr}
\affiliation{Department of Physics and Astronomy, Uppsala University, 751 20 Uppsala, Sweden}

\author{Claudio Serpico}
\affiliation{DIETI, University of Naples Federico II, 80125 Naples, Italy}

\author{Matthias C. Hoffmann}
\affiliation{SLAC National Accelerator Laboratory, Menlo Park, CA 94025, USA}

\author{Stefano Bonetti}
\email{stefano.bonetti@fysik.su.se}
\affiliation{Department of Physics, Stockholm University, 106 91 Stockholm, Sweden}
\affiliation{Department of Molecular Sciences and Nanosystems, Ca' Foscari University of Venice, 30172 Venezia Mestre, Italy}

\date{\today}

\begin{abstract}
We present a comprehensive experimental and numerical study of magnetization dynamics in a thin metallic film triggered by single-cycle terahertz pulses of $\sim20$ MV/m electric field amplitude and $\sim1$ ps duration. The experimental dynamics is probed using the femtosecond magneto-optical Kerr effect, and it is reproduced numerically using macrospin simulations. The magnetization dynamics can be decomposed in three distinct processes: a coherent precession of the magnetization around the terahertz magnetic field, an ultrafast demagnetization that suddenly changes the anisotropy of the film, and a uniform precession around the equilibrium effective field that is relaxed on the nanosecond time scale, consistent with a Gilbert damping process. Macrospin simulations quantitatively reproduce the observed dynamics, and allow us to predict that novel nonlinear magnetization dynamics regimes can be attained with existing table-top terahertz sources.
\end{abstract}

\pacs{Valid PACS appear here}

\maketitle

Since Faraday's original experiment~\cite{Faraday01011846} and until two decades ago, the interaction between magnetism and light has been mostly considered in a unidirectional way, in which changes to the magnetic properties of a material cause a modification in some macroscopic observable of the electromagnetic radiation, such as polarization state or intensity. However, the pioneering experiment of Beaurepaire \textit{et al.}~\cite{beaurepaire1996ultrafast}, where femtosecond optical pulses were shown to quench the magnetization of a thin-film ferromagnet on the sub-picoseconds time scales, demonstrated that intense laser fields can conversely be used to control magnetic properties, and the field of ultrafast magnetism was born. Large research efforts are nowadays devoted to the attempt of achieving full and deterministic control of magnetism using ultrafast laser pulses~\cite{koopmans2000ultrafast,vanKampen02,koopmans2005unifying,stanciu2007all,malinowski2008control,Koopmans:NatureMaterials:2009,kampfrath2011coherent,Kubacka1333}, a fundamentally difficult problem that could greatly affect the speed and efficiency of data storage~\cite{Walowski16}.

Recently, it has been shown that not only femtosecond laser pulses, but also intense single-cycle terahertz (THz) pulses \cite{hoffmann2011coherent} can be used to manipulate the magnetic order at ultrafast time scales in different classes of materials ~\cite{Nakajima10,Yamaguchi10,Kim14,Bonetti16,Shalaby16, schlauderer2019temporal}. The main peculiarity of this type of radiation, compared with more conventional femtosecond infrared pulses, is that the interaction with the spins occurs not only through the overall energy deposited by the radiation in the electronic system, but also through the Zeeman torque caused by the magnetic field component of the intense THz pulse. This is a more direct and efficient way of controlling the magnetization, and to achieve the fastest possible reversal~\cite{tudosa2004ultimate,gamble2009electric}. However, an accurate description of the magnetization dynamics triggered by strong THz pulses is still missing.

In this Letter, we present a combined experimental and numerical study of the magnetization dynamics triggered by linearly polarized single-cycle THz pulses with peak electric (magnetic) fields up to 20 MV/m (67 mT). We investigate not only the fast time scales that are comparable to the THz pulse duration ($\sim1$ ps), but also the nanosecond regime, where ferromagnetic resonance (FMR) oscillations are observed. Moreover, we write an explicit form of the Landau-Lifshitz-Gilbert (LLG) equation~\cite{Landau35,Gilbert04} suitable to analyze terahertz-driven dynamics, that we use to predict yet-unexplored nonlinear magnetization dynamics regimes uniquely achievable with this type of excitation mechanism.

\begin{figure}[ht!]
\centering\includegraphics[width=\columnwidth]{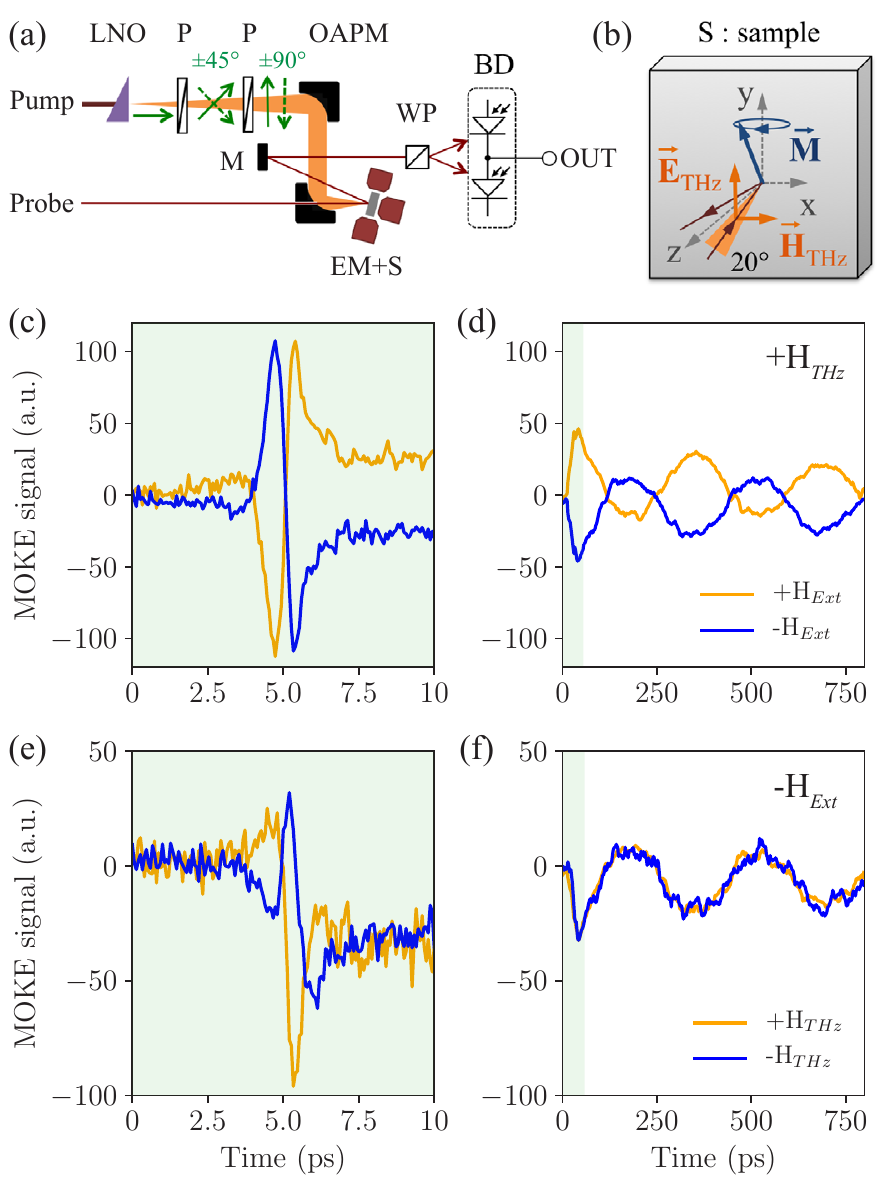}
\caption{(Color online) (a) Schematic drawing of the experimental setup: BD - Balanced detection using two photodiodes and a lock-in amplifier, WP - Wollaston prism, EM + S - Electromagnet  with out-of-plane field and sample, OAPM - Off-axis parabolic mirror, P - Wire grid polarizer, LNO - Lithium niobate, and M - Mirror. (b) Sample geometry: The electric field component of the THz pulse at the sample position is oriented parallel to the $y$-axis direction, $\textrm{\bf E}_{\textrm{THz}}\parallel \textbf{y}$, and the magnetic field component parallel to the $x$-axis direction, $\textrm{\bf H}_{\textrm{THz}}\parallel \textbf{x}$. A static magnetic field $\textrm{\bf H}_{\textrm{Ext}}$ is applied along the $z$-axis direction. (c-f) Experimental MOKE data  showing the influence of reversing the external magnetic field ($\pm\textrm{\bf H}_{\textrm{Ext}}$, $\textrm{\bf H}_{\textrm{THz}}=$ const.) on the THz-induced demagnetization (c) and on the FMR oscillations (d). The influence of reversing the THz magnetic field pulse ($\pm\textrm{\bf H}_{\textrm{THz}}$, $\textrm{\bf H}_{\textrm{Ext.}}=$ const.) on the THz-induced demagnetization and on the FMR oscillations is shown in (e) and (f), respectively. (The shaded green area is a guide to the eye.)}
\label{F1}
\end{figure}

\emph{Experimental details.}
Room temperature experimental data is obtained from a time-resolved pump-probe method utilizing the magneto-optical Kerr effect (MOKE), Refs.~\cite{Freeman96,Hiebert97}. A sketch of the experimental setup is presented in Fig.~\ref{F1} (a). Strong THz radiation is generated via optical rectification of 4 mJ, 800 nm, 100 fs pulses from a 1 kHz regenerative amplifier in a lithium niobate (\ce{LiNbO3}) crystal, utilizing the tilted-pulse-front method~\cite{hoffmann2011intense}. In contrary to the indirect (thermal) coupling present in visible- and near-infrared light-matter interaction, THz radiation can directly couple to the spin system via magnetic dipole interaction (Zeeman interaction)~\cite{Hirori16}. In this respect, a fundamental aspect is the orientation of the THz polarization, which is controlled using a set of two wire grid polarizers, one variably oriented at $\pm$45$^{\circ}$ and a second one fixed to +90$^{\circ}$ (or -90$^{\circ}$) with respect to the original polarization direction of $\textrm{\bf E}_{\textrm{THz} }$. As depicted in Fig.~\ref{F1} (b), the magnetic field component of the THz pulse $\textrm{\bf H}_{\textrm{THz} }$ is fixed along the $x$-axis direction and is therefore flipped by 180$^{\circ}$ by rotating the first polarizer. An amorphous \ce{CoFeB} sample (\ce{Al2O3} (1.8nm)/\ce{Co40Fe40B20} (5nm)/\ce{Al2O3} (10nm)/Si substrate) is placed either in the gap of a $\pm$200 mT electromagnet or on top of a 0.45 T permanent magnet. In both cases, the orientation of the externally applied field $\textrm{\bf H}_{\textrm{Ext} }$ is along the $z$-axis direction, i.e.\ out of plane with respect to the sample surface. However, a small component of this external bias field lying in the sample plane parallel to the $y$-axis direction has to be taken into account, due to a systematic (but reproducible) small misalignment in positioning the sample. The THz pump beam, with a spot size \diameter $\approx 1$ mm (FWHM), and the 800 nm probe beam, with a spot size \diameter $\approx 200$ $\mu$m (FWHM), overlap spatially on the sample surface in the center of the electromagnet gap. Being close to normal incidence, the MOKE signal is proportional to the out-of-plane component ${\rm M}_{\rm z}$ of the magnetization, i.e.\ polar MOKE geometry \cite{qiu2000surface}. The probe beam reflected from the sample surface is then analyzed using a Wollaston prism and two balanced photo-diodes, following an all-optical detection scheme~\cite{vanKampen02}. 

\emph{Results and discussion.}
The experimental data demonstrating THz-induced demagnetization and the magnetic field response of the spin dynamics is shown in Fig.~\ref{F1} (c-f) when ${\mu_0}\textrm{H}_{\textrm{Ext}}=185\ {\rm mT}$. For short timescales on the order of the THz pump pulse, $\tau \sim$ 1 ps, the polar MOKE is sensitive to the coherent response of the magnetization, i.e.\ its precession around the THz magnetic field, as shown in Fig.~\ref{F1} (c)+(e). Within $\tau \sim$ 100 fs after time zero (${\rm t_0}\sim$ 5 ps) a sudden demagnetization step of the order of 0.1-0.2\% of the total magnetization vector is observed. The demagnetization step is followed by a 'fast' relaxation process, $\tau \sim$ 1 ps, and subsequently by a 'slow' recovery of the magnetization on a longer timescale, $\tau \sim$ 100 ps. During and after magnetization recovery, a relaxation precession (corresponding to the FMR) is superimposed, see Fig.~\ref{F1} (d)+(f). The effect of reversing the externally applied magnetic field $\textrm{\bf H}_{\textrm{Ext}}$ on the MOKE measurements is illustrated in Fig.~\ref{F1} (c)+(d). This data shows that all the different processes just identified (demagnetization, coherent magnetization response in the range ${\rm t_0}<{\rm t}\lesssim{\rm t_0} + {\rm 2}$ ps, and FMR response) are indeed magnetic, as they all reverse their sign upon reversal of the sign of the bias magnetic field. Fig.~\ref{F1} (e)+(f) instead depict the effect of reversing the THz field polarity, while keeping the externally applied field constant. This data illustrate the symmetry of the magnetic response with respect to the THz magnetic field. The demagnetization and FMR response, whose amplitude scales quadratically with the THz magnetic field $\textrm{\bf H}_{\textrm{THz}}$, remain unchanged upon reversal of the terahertz magnetic field, while the coherent magnetization response, linear in $\textrm{\bf H}_{\textrm{THz}}$, reverses its sign.

\begin{figure}[t!]
\centering\includegraphics[width=0.9\columnwidth]{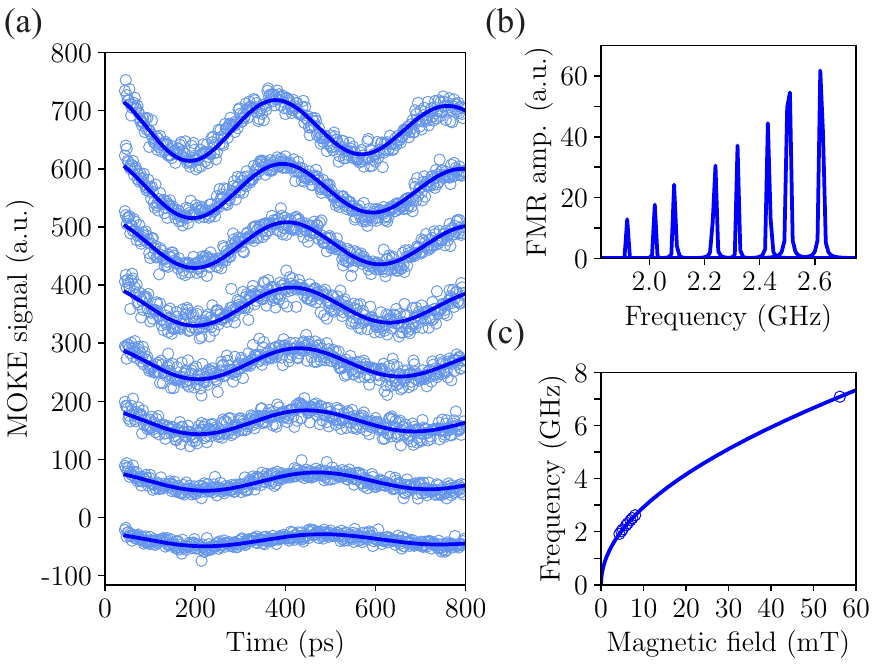}
\caption{(Color online) Experimental data showing the magnetic field dependence of the FMR. (a) MOKE signal for different out-of-plane fields ${\mu_0}\textrm{H}_{\textrm{Ext}}$ - 20, 50, 75, 100, 125, 150, 175 and 185 mT in ascending order. The solid line is a damped sine function fitted to the experimental data. Both fit and data curve are displaced on the y axis by a constant offset. (b) Fourier transformation of the data shown in (a) vs. FMR amplitude. (c) FMR frequency as a function of the in-plane component of $\textrm{\bf H}_{\textrm{Ext}}$. The solid line corresponds to Kittel equation, Eq.~\eqref{eq:kittel}.}
\label{F2}
\end{figure}
The dependence of the THz-induced FMR response on the magnitude of the magnetic bias field ($\textrm{\bf H}_{\textrm{Ext}}$) is shown in Fig.~\ref{F2} (a). Oscillation amplitude and resonance frequency are summarized in Fig.~\ref{F2} (b). The oscillation amplitude is obtained from fitting a damped sinusoidal function to the data shown in Fig.~\ref{F2} (a), and the resonance frequency follows from subsequent Fourier transformation of that fitting function. The relationship between magnetic resonance field ${\bf H}_{\rm r}$ and FMR frequency $f_{\rm FMR}$ can be described by the phenomenological Kittel equation~\cite{Kittel48}
\begin{equation}
f_{\rm FMR}=\dfrac{\gamma\mu_0}{2\pi}\sqrt{{\bf H}_{\rm r}({\bf H}_{\rm r}+{\bf M}_{\rm eff})}
\label{eq:kittel}
\end{equation}
with gyromagnetic ratio $\gamma$ and effective magnetization ${\bf M}_{\rm eff} = {\bf M}_{\rm S} - {\bf H}^{\perp}_{\rm K} $, where $ {\bf H}^{\perp}_{\rm K} $ is the perpendicular anisotropy field. From room-temperature magnetization measurements (see the Supplemental Material), a saturation magnetization $\mu_0{\rm M}_{\rm S}=$ 1.84 T and an anisotropy field $\mu_0{\rm H}^{\perp}_{\rm K}=$ 0.76 T are obtained. The value of ${\rm H}_{\rm r}$ represents the small in-plane component of $\textrm{\bf H}_{\textrm{Ext}}$ (i.e.\ along the $y$-axis direction), which can be inferred from the measurements in Fig.~\ref{F2} (a) in order to get good agreement with the experimental data, as shown in Fig.~\ref{F2} (c). 

\begin{figure}[t!]
\centering\includegraphics[width=1\columnwidth]{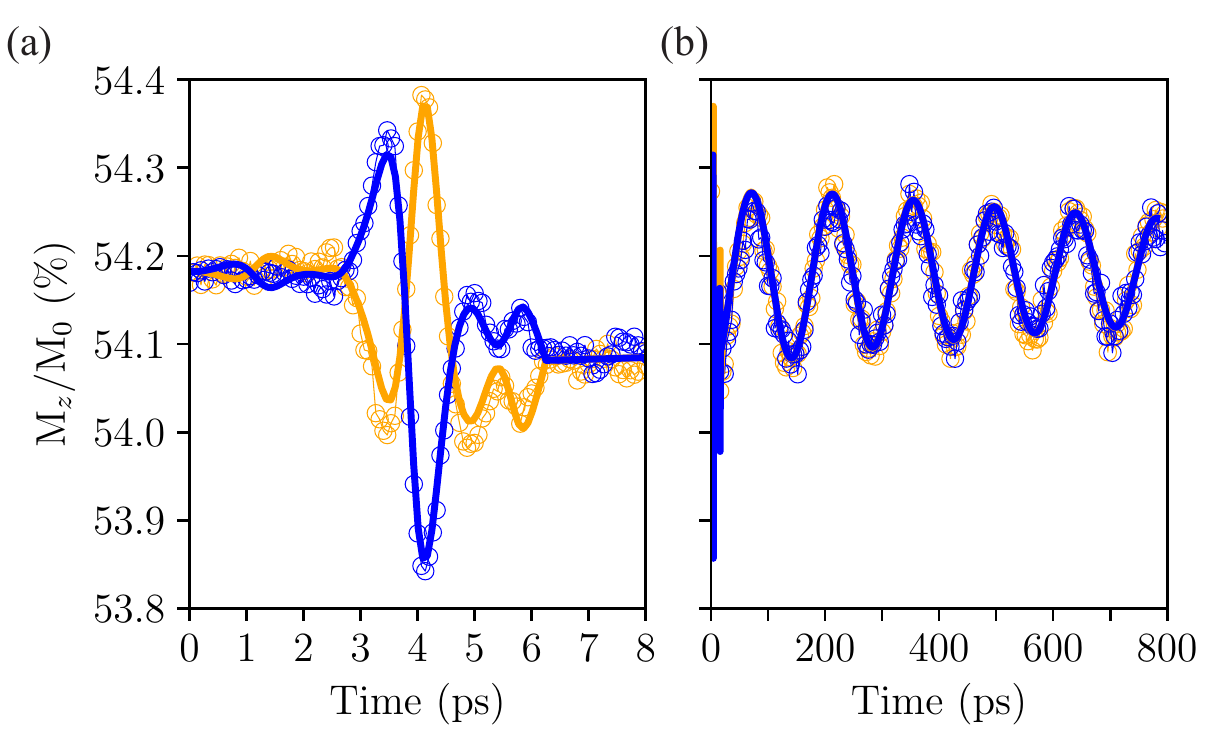}
	\caption{(Color online) Comparison of experimental data (open markers) and macrospin simulations (solid lines). (a) THz-induced demagnetization for an external applied field of $\mu_0 \textrm{H}_{\rm Ext}$ $\sim$ 450 mT and a THz pulse of $\textrm{E}_{\textrm{THz}}$ =  18 MV/m, ($\mu_0 \textrm{H}_{\textrm{THz}}\sim$ 60 mT). The blue markers/lines correspond to $+\textrm{\bf H}_{\textrm{THz}}$, and the orange markers/lines to $-\textrm{\bf H}_{\textrm{THz}}$. (b) Experimental data and macrospin simulations under similar conditions as (a) showing FMR on a longer timescale.}
\label{F3}
\end{figure}

\emph{Numerical modelling.}
A detailed picture of the THz-induced demagnetization and FMR response for an external magnetic field generated by a 0.45 T permanent magnet and a THz pulse of $\textrm{E}_{\textrm{THz}}$ = 18 MV/m ($\mu_0 \textrm{H}_{\textrm{THz}}\sim$ 60 mT) is presented in Fig.~\ref{F3}. Experimental data has been recorded both on a short timescale after arrival of the pump pulse showing coherent precession and demagnetization, Fig.~\ref{F3} (a), and on long timescales showing magnetization recovery and FMR at a frequency of about 7.2 GHz, Fig.~\ref{F3} (b). By knowing the FMR frequency, the actual applied magnetic field can be deduced from the data shown in Fig.~\ref{F2} (a) and Eq.~\eqref{eq:kittel}, as described in the previous paragraph. Following this approach, we find a value of 450 mT for the out-of-plane component and 56 mT for the in-plane component of $\textrm{\bf H}_{\textrm{Ext}}$, corresponding to the right-most point in Fig.~\ref{F2} (c). 

This complete set of good-quality data has been used for validating all the assumptions considered in our macrospin simulations, whose results are shown as solid lines in Fig.~\ref{F3}. Indeed, magnetization dynamics of a uniform ferromagnetic material can be modeled and understood from macrospin simulations where the description of the magnetic state is simplified by a single-domain approximation. The phenomenological LLG equation~\cite{Gilbert04} can be used to obtain a first approximate description of the magnetization continuum precession and relaxation. Since the LLG equation preserves the length of the magnetization vector (${|\textbf{M}|}=$ const.)~\cite{Horley11}, it does not account for laser-induced demagnetization and it ignores relevant physical phenomena such as scattering effects~\cite{Walowski16}. A possibility to account for optically- or thermally-induced demagnetization is the use of the Landau-Lifshitz-Bloch (LLB) equation~\cite{Garanin90}, which has been shown to describe ultrafast demagnetization processes~\cite{Atxitia17}. However, the parameters involved in the LLB equation depend on temperature, and a description of their temporal evolution due to the interaction with ultrafast pulses is needed. Such a description is often provided using a two-temperature model~\cite{Koopmans:NatureMaterials:2009,Atxitia17}, which then has to be coupled to the LLB equation. However, such a two-temperature model relies on several phenomenological material parameters, and little insight on the physics is gained by this approach, in particular at longer time scales. Since here we are interested in these time scales, we simplify the problem by just assuming a phenomenological LLG equation with non-constant magnetization magnitude, a physical quantity which can be readily measured.

Our version of the LLG equation reads (see the Supplemental Material)
\begin{equation}
\frac{d\mathbf{M}}{dt} = -\gamma' \big[ \mathbf{M} \times \mathbf{H_{e}} +\frac{ \alpha}{{\rm {M_S}}} (\mathbf{M} \times (\mathbf{M} \times \mathbf{H_{e}})) \big]- c(\mathbf{M},\mathbf{H_{e}})\mathbf{M},
\label{eq:llg}
 \end{equation}
\noindent with $\gamma'=\gamma/(1+\alpha^2)$, where the gyromagnetic ratio is $|\gamma/2\pi| \approx$ 28.025 GHz/T, and $\mathbf{M}=\rm M \mathbf{m}$, where ${\rm M}(t)$ is the magnetization amplitude and $\mathbf{m}(t)$ the unit vector. The effective magnetic field $\mathbf{H_{e}}$ includes the external field $\mathbf{H_{\rm Ext}}$, THz field $ \mathbf{H_{\rm THz}}$ and demagnetization field $ \mathbf{H_{\rm D}}$. The Gilbert damping $\alpha=0.01$ was independently measured with conventional FMR spectroscopy, as shown in the Supplemental Material. In Eq.\ \eqref{eq:llg}, the first and second terms on the right-hand side describe coherent spin precession and the macroscopic spin relaxation, respectively, whereas the third term describes the fast demagnetization along the $\mathbf{m}$ direction controlled by the function $c(\mathbf{M},\mathbf{H_{e}})$. A non-constant magnetization magnitude is also needed for reproducing the observed FMR oscillation, as discussed in the Supplemental Material.

Hence, we propose a semi-empirical approach to identify $c(\mathbf{M},\mathbf{H_{e}})$ and, consequently, the time evolution of the magnetization vector length ${\rm M}(t)$ from experimental demagnetization data. For this purpose, the change in ${\rm M}(t)$ as a function of time is calculated from the cumulative integral of the incident THz pulse~\cite{Bonetti16}:
\begin{equation}
    \Delta {\rm M}(t)\propto {\rm e}^{-t/\tau_R}\int_{-\infty}^t \mathrm{H_{THz}}(\xi)^2d\xi,
    \label{eq:demag}
\end{equation}
where the exponential term describes the recovery of the magnetization on a timescale of $\sim$100 ps for \ce{CoFeB}. The shape of the incident THz pulse (in the time domain) is proportional to $\mathbf{E_{\rm THz}}(t)$, which can be measured via electro-optical sampling in GaP~\cite{Hoffmann:11}. We notice that the THz magnetic field $\mathbf{H_{\rm THz}}(t)$ inside the material, to be used in Eq.\ \eqref{eq:demag}, can be derived from the measured free-space $\mathbf{E_{\rm THz}}(t)$ after taking into account the whole sample stack, e.g.\ by means of the transfer matrix method~\cite{bornwolf}. Eventually, the function $c(\mathbf{M},\mathbf{H_{e}})$ is obtained from Eq.\ \eqref{eq:demag} after re-scaling it with experimentally obtained demagnetization values (see Ref.~\cite{Bonetti16} for more details).

After determining $c(\mathbf{M},\mathbf{H_{e}})$, all the terms in Eq.\ \eqref{eq:llg} are known, and the equation can be used to compute the dynamics of the magnetization orientation in terms of the spherical angles $\theta(t),\ \phi(t)$. Indeed, Eq.\ \eqref{eq:llg} is numerically solved in spherical coordinates $({\rm M},\theta,\phi)$ (see the Supplemental Material for details) using a custom-made Python solver based on the 4th-order Runge-Kutta method, which provides a simple implementation and good accuracy of the numerical solution~\cite{Horley11}.

It is now interesting to explore the expected response of the magnetization to THz fields with strength larger than the ones considered in this work, but nowadays accessible with table-top sources. For simplicity and generality, only the free-space value of the THz peak electric field is reported in the below discussion. In case of THz fields smaller than ${\rm E}_{\rm THz}\sim$ 100 ${\rm MV/m}$, the demagnetization follows the square of the amplitude of the THz field, see Ref.~\cite{Bonetti16}. Lacking detailed experimental data, it is reasonable to assume that THz-induced demagnetization for higher THz peak fields (${\rm E}_{\rm THz}>$ 100 ${\rm MV/m}$) can be described by the positive section of an error function, allowing for a quadratic behavior for small demagnetization and a saturation for large demagnetization approaching 100\%~\cite{Shalaby16}. From our experimental data, we derive a functional description of the demagnetization as a function of the THz field such as Demag $ ={\rm f(E)}$ = ${\rm erf}(A\cdot{\rm E}^2)$, with THz peak field ${\rm E}$ and fitting parameter $A\approx6.0\cdot10^{-6}$ $\mathrm{m^2\ V^{-2}}$. (See the Supplemental Material for further details.)
\begin{figure}[t!]
\centering
\includegraphics[width=\columnwidth]{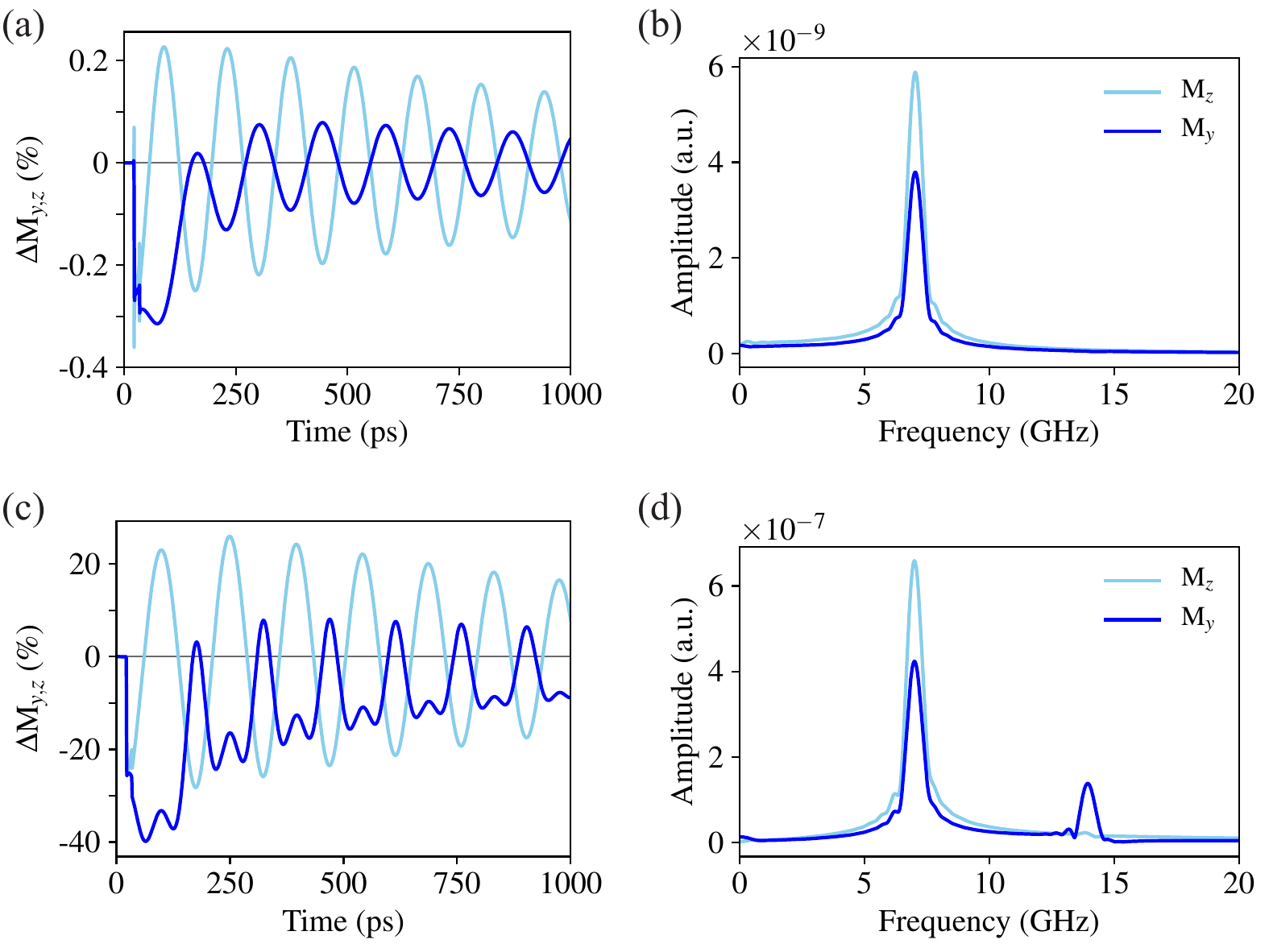}
\caption{(Color online) Comparison of macrospin simulation data for moderate (20 MV/m) and high (200 MV/m) THz-peak-field stimulus. The external magnetic field is set to $\mu_0 \textrm{H}_{\textrm{Ext}}$ = 450 mT, i.e.\ replicating the experimental conditions from Fig.~\ref{F3}.  Panel (a) shows the relative change of the ${\rm M}_{\rm y}$ (blue line) and ${\rm M}_{\rm z}$ (light blue line) magnetization component for a ${\rm E}_{\rm THz}$ peak field of 20 MV/m and (b) shows the corresponding Fourier spectrum. (c) and (d) show the simulation data and  corresponding Fourier spectrum for an ${\rm E}_{\rm THz}$ peak field of 200 MV/m, respectively.}
\label{F4}
\end{figure}

With this assumption, the macrospin simulation results for THz fields ${\rm E}_{\rm THz}=$ 20 ${\rm MV/m}$ and ${\rm E}_{\rm THz}=$ 200 ${\rm MV/m}$ are presented in Fig.~\ref{F4} (a-b) and Fig.~\ref{F4} (c-d), respectively. For ${\rm E}_{\rm THz}=$ 200 ${\rm MV/m}$, a clear nonlinear response of the magnetization to the THz field is found, illustrated by the second harmonic oscillation in the ${\rm M}_y$ component of the magnetization. The simulated THz-induced demagnetization for ${\rm E}_{\rm THz}=$ 200 ${\rm MV/m}$ is on the order of $\Delta{\rm M}_{\rm z} \sim$ 20\%. In Fig.~\ref{F4} (b)+(d), the Fourier spectrum of the FMR oscillation for the ${\rm M}_y$ and ${\rm M}_z$ components of the magnetization at THz pump peak fields of 20 MV/m and 200 MV/m are depicted. The Fourier data of the 200 MV/m simulation shown in Fig.~\ref{F4} (c) clearly shows a second harmonic peak at $\sim$ 14 GHz, present for ${\rm M}_y$ but not for ${\rm M}_z$.
A similar behavior was observed recently by performing FMR spectroscopy of thin films irradiated with femtosecond optical pulses inducing either ultrafast demagnetization \cite{Capua16}, by exciting acoustic waves \cite{Chang17}, and by two-dimensional THz magnetic resonance spectroscopy of antiferromagnets \cite{lu2017coherent}. In our case, the high-harmonic generation process is solely driven by the large amplitude of the terahertz magnetic field that is completely off-resonant with the uniform precession mode. This would allow for exploring purely magnetic dynamics in regimes that are not accessible with conventional FMR spectroscopic techniques, where high-amplitude dynamics are prevented by the occurrence of so-called Suhl's instabilities, i.e.\ non-uniform excitations degenerate in energy with the uniform mode. Such non-resonant, high THz magnetic fields are within the capabilities of recently developed table-top THz sources \cite{shalaby2018coherent}, and can also be generated in the near-field using metamaterial structures as described by Refs.~\cite{Mukai16,polley2018THz-driven,polley2018terahertz}.

In summary, we investigated magnetization dynamics induced by moderate THz electromagnetic fields in amorphous \ce{CoFeB}, in particular the ferromagnetic resonance response as a function of applied bias and THz magnetic fields. We demonstrate that semi-empirical macrospin simulations, i.e.\ solving the Landau-Lifshitz-Gilbert equation with a non-constant magnitude of the magnetization vector to incorporate THz-induced demagnetization effect, are able to describe all the details of the experimental results to a good accuracy. Existing models of terahertz spin dynamics and spin pumping would need to be extended to include the evidence presented here \cite{bocklage2016model,bocklage2017coherent}. Starting from simulations describing experimental data for THz-induced demagnetization, we extrapolate that THz fields one order of magnitude larger drive the magnetization into a nonlinear regime. Indeed, macrospin simulations with THz fields on the order ${\rm E}_{\rm THz}\sim$ 200 ${\rm MV/m}$ ($\mu_0 \textrm{H}_{\textrm{THz}}\sim$ 670 mT) predict a significant demagnetization of $\Delta{\rm M}_{\rm z} \sim$ 20\%, and a marked nonlinear behavior, apparent from second harmonic generation of the uniform precessional mode. We anticipate that our results will stimulate further theoretical and experimental investigations of nonlinear spin dynamics in the ultrafast regime. 

M.H.\ gratefully acknowledges support from the Swedish Research Council grant E0635001, and the Marie Sk\l{}odowska Curie Actions, Cofund, Project INCA 600398s. The work of M.d'A.\ was carried out within the Program for the Support of Individual Research 2017 by University of Naples Parthenope. M.C.H.\ is supported by the U.S.\ Department of Energy, Office of Science, Office of Basic Energy Sciences, under Award No.\ 2015-SLAC-100238-Funding. M.P.\ and S.B.\ acknowledge support from the European Research Council, Starting Grant 715452 ``MAGNETIC-SPEED-LIMIT''.

\bibliography{THz_lit}

\end{document}